\documentclass{onecolartcl}
\usepackage{graphicx} 

\title{PyDislocDyn: A Python code for calculating dislocation drag and other crystal properties}
\author{Daniel N. Blaschke}
\date{November 4, 2025}

\begin{document}

\maketitle

\thispagestyle{empty}
\begin{center}
\renewcommand{\thefootnote}{\fnsymbol{footnote}}
\vspace{-0.5cm}
Los Alamos National Laboratory, Los Alamos, NM, 87545, USA
\\[0.2cm]
\ttfamily{E-mail: dblaschke@lanl.gov}
\\{\small LA-UR-25-28917}
\end{center}


\section*{Summary}
PyDislocDyn is a suite of python programs designed to perform various calculations for basic research in dislocation dynamics in metals with various crystal symmetries in the continuum limit.
In particular, one of its main purposes is to calculate dislocation drag from phonon wind.
Additional features include the averaging of elastic constants for polycrystals, the calculation of the dislocation field including its limiting velocities, and the calculation of dislocation self-energy and line tension.

\section*{Introduction}
Line defects in crystals, known as dislocations, play an important role in accommodating plastic deformation by gliding through the crystal \cite{Hansen:2013,Lloyd:2014,Luscher:2016,Austin:2018,Blaschke:2021impact,Zuanetti:2021}.
A good understanding of their properties, including how dislocations interact with obstacles (like grain boundaries, impurities, and other defects), is thus key to understanding material strength.
At high deformation rates, gliding dislocations experience an opposing drag force due to phonon scattering, an effect known as phonon wind \cite{Weertman:1980,Nadgornyi:1988,Alshits:1992,Gurrutxaga:2020}.
In a series of papers summarized in the review article of Ref. \cite{Alshits:1992}, Alshits and co-workers developed a first-principles theory of dislocation drag in the continuum limit of an isotropic crystal, assuming small gliding velocities (i.e. a few percent of the transverse sound speed).
A generalization of the phonon wind theory to large dislocation gliding velocities (up to the transverse sound speed) was later developed in Refs. \cite{Blaschke:BpaperRpt,Blaschke:2019Bpap}, and subsequently further generalized to anisotropic crystals in Ref. \cite{Blaschke:2018anis}.
Simplified functional forms of dislocation drag (in some cases inspired by this theory \cite{Blaschke:2021impact}) are used in single crystal plasticity simulations \cite{Lloyd:2014JMPS,Luscher:2016,Austin:2018,Zuanetti:2021,Ye:2023a,Manukhina:2024} 
as well as discrete dislocation dynamics (DDD) simulations \cite{Gurrutxaga:2013,Bertin:2015,Cho:2015,Kim:2020,Akhondzadeh:2023,Tak:2023,Bertin:2024}.

Dislocation theory also predicts limiting velocities for dislocation glide where the elastic strain energy of the moving dislocation (and thereby also the drag coefficient from phonon wind) diverges \cite{Teutonico:1961,Weertman:1962,VanHull:1962,Teutonico:1962,Blaschke:2021vcrit}.
This prediction, however, has its roots in the simplifying assumption of a ``perfect'' dislocation, which neglects the fact that dislocations have a finite core size in a real crystal.
Even in the continuum theory, these velocities can be overcome in principle when the dislocation core is taken into account in a regularising fashion, see \cite{Markenscoff:2008,Pellegrini:2018,Pellegrini:2020}.
Thus within real crystals, the ``limiting velocities'' can be seen as dislocation velocities that are hard, but not necessarily impossible, to overcome.
In fact, a number of molecular dynamics (MD) simulations of various cubic and hexagonal close-packed (hcp) metals (see e.g. \cite{Gumbsch:1999,Olmsted:2005,Tsuzuki:2008,Oren:2017,Peng:2019,Dang:2022Mg,Duong:2023a,Jones:2025MSMSE} and references therein),
as well as one experiment \cite{Katagiri:2023} on diamond, have predicted that dislocations can also glide at transonic or even supersonic speeds (i.e. above their highest limiting velocity).
Due to technological limitations, tracking high speed dislocations in metals in real time remains challenging.
Therefore, no transonic or supersonic dislocations have been directly observed in metals to date and the according MD results remain unconfirmed experimentally.

Apart from questions regarding the highest dislocation gliding velocities, mesoscale simulations in general also suffer from high uncertainties in modeling dislocation density evolution at high strain rates.
For recent advances of the latter in face-centered cubic (fcc) metals, see \cite{Hunter:2022,Larkin:2022} and references therein.
Dislocation density evolution, however, is beyond the scope of our present code, PyDislocDyn, which focuses on properties of single dislocations.

\section*{Statement of need}

Large-scale simulations of how metals (and other crystals) deform, such as DDD simulations \cite{Gurrutxaga:2013,Bertin:2015,Cho:2015,Kim:2020,Akhondzadeh:2023,Tak:2023,Bertin:2024} and crystal plasticity simulations \cite{Lloyd:2014JMPS,Luscher:2016,Austin:2018,Zuanetti:2021,Ye:2023a,Manukhina:2024}, can increase their fidelity by including dislocation dynamics properties (drag coefficient, limiting velocities, etc.).
PyDislocDyn can aid researchers in those endeavors.
In particular, one of the main purposes of the open source code PyDislocDyn \cite{pydislocdyn} is to provide a reference implementation of the first-principles theory of dislocation drag developed in Refs. \cite{Blaschke:2019Bpap,Blaschke:2018anis}.
As one of the ingredients of phonon wind theory, the dislocation displacement gradient field in the continuum limit needs to be computed.
PyDislocDyn makes use of the computationally efficient ``integral method'' for steady-state dislocations developed by Lothe and co-workers, which builds upon the earlier ``sextic'' formalism of Stroh and others.
A review of this method can be found in the excellent article of Bacon et al. \cite{Bacon:1980} as well as the textbook of Hirth and Lothe \cite{Hirth:1982}.

Accounting for accelerating dislocations in phonon wind calculations would be too costly, computationally, given the expected subleading effect, see \cite{Markenscoff:1987,Blaschke:2020acc,Blaschke:2023acc} and references therein.
Nonetheless, the displacement gradient fields of accelerating pure edge and screw dislocations for those slip systems featuring a reflection symmetry can be calculated separately within the code.
While dislocations of arbitrary character angle are modeled according to the crystal symmetry (bcc, fcc, hcp, etc.) and slip system, the phonon spectrum is kept isotropic for simplicity \cite{Blaschke:2018anis}.

Additionally, PyDislocDyn includes an implementation of the limiting velocities of dislocation glide following the recipe outlined in the review article \cite{Blaschke:2021vcrit} in order to limit dislocation drag calculations to their range of validity on the one hand, but also to enable users to determine those limiting velocities in an accessible way for other purposes, such as interpreting molecular dynamics simulation results of dislocation glide \cite{Blaschke:2020MD,Dang:2022Mg,Jones:2025MSMSE}.

\section*{Features}

Its main features can be summarized as:
\begin{itemize}
\item The code can compute isotropic averages of elastic constants using various methods, such as Voigt-Reuss-Hill averaging and Kr{\"o}ner's method, see \cite{Blaschke:2017Poly}.
\item It can compute the steady-state displacement (gradient) field of a dislocation for arbitrary character angle in an arbitrary crystal geometry \cite{Bacon:1980,Hirth:1982}; a generalization to accelerating dislocations has been implemented for pure screw dislocations \cite{Blaschke:2020acc} and pure edge dislocations \cite{Blaschke:2023acc}.
\item We compute theoretical limiting velocities of dislocations in arbitrary slip systems and with arbitrary character angle. \cite{Blaschke:2021vcrit,Blaschke:2020MD,Teutonico:1961,Barnett:1973b}.
\item The code can compute the elastic strain energy and line tension for any dislocation \cite{Blaschke:2017lten}.
\item And most importantly, PyDislocDyn can compute the dislocation drag coefficient $B$ from phonon wind for any dislocation at any velocity \cite{Blaschke:BpaperRpt,Blaschke:2018anis,Blaschke:2019fits,Blaschke:2019Bpap} or stress \cite{Blaschke:2019a,Blaschke:2021impact,Blaschke:2021temperature}.
Given appropriate elastic constants, dislocation drag can in principle be computed for any temperature or pressure with PyDislocDyn as elucidated in Ref. \cite{Blaschke:2021temperature}.
While calculating elastic constants is beyond the scope of PyDislocDyn, we do provide some tools to determine which types of deformations are sensitive to which elastic constants.
This information can then be used as a starting point to computing elastic constants using other third party tools  \cite{Blaschke:2021temperature,Gu:2019}.
\end{itemize}
A number of additional tools for various crystal operations are available as well.
In particular:
computing sound speeds for any direction within the crystal, computing Rayleigh wave speeds \cite{Barnett:1973b}, converting various tensors to and from Voigt notation, converting between different sets of isotropic elastic constants, computing measures of anisotropy \cite{Kube:2016anis}, determining 'radiation-free' dislocation velocities \cite{Gao:1999,Blaschke:2023rad}, converting Miller indices to Cartesian coordinates, computing the elastic compliance tensors from the elastic constant tensors, converting the drag coefficient $B(v)$ to $B(\sigma)$ (i.e. a function of resolved shear stress instead of a function of velocity), and various visualization functions for the dislocation field and its drag coefficient.
Many functions also support Sympy manipulations.
For example, Voigt-Reuss-Hill averages of elastic constants can be computed symbolically as well as numerically.
For further details on functionality, we refer the reader to the code manual and the Jupyter notebook containing examples, which are both included in the code distribution available on Github \cite{pydislocdyn}.

\section*{Acknowledgements}

Support from the Materials project within the Physics and Engineering Models (PEM) Subprogram element of the Advanced Simulation and Computing (ASC) Program at Los Alamos National Laboratory (LANL)is gratefully acknowledged.
LANL, an affirmative action/equal opportunity employer, is operated by Triad National Security, LLC, for the National Nuclear Security Administration of the U.S. Department of Energy under contract 89233218NCA000001.

\bibliographystyle{utphys-custom}
\bibliography{joss_pydislocdyn}
\end{document}